# MEASURING ORBIT RESPONSES WITH OSCILLATING TRAJECTORIES IN THE FERMILAB LINAC*

A. Shemyakin†, A. Pathak, R. Sharankova
Fermilab, Batavia, IL 60510, USA

*Abstract*

Recording changes in beam transverse positions reported by Beam Position Monitors (BPMs) in response to a beam deflection by an upstream dipole corrector (orbit response) is a powerful tool for analysis of accelerator optics and assisting with machine tuning. In the Fermilab Linac, orbit responses are recorded by oscillating the currents of up to 19 correctors, providing faster, drift-resistant measurements through frequency-domain analysis. This report describes the technique, including error estimations and consistency checks and shows an example of the measurements.

## INTRODUCTION

The linear properties of transverse accelerator optics can be characterized using Orbit Response Matrix (ORM) that contains responses of BPMs to changes in currents of dipole correctors ([1],[2]). Initially started in the circular accelerators, similar techniques are being applied now to linear accelerators as well [3]. A powerful way to record the ORM is to use sinusoidal excitation of the correctors. It is used in the rings (for example, [4], [5]) but commonly not in the linacs. Here we are describing the method (mostly following Ref. [6]) and details of its implementation at Fermilab 400 MeV Linac and show a corresponding example.

## METHOD

In a beam line or linear accelerator with uncoupled linear optics, the change of the beam position $\Delta x_j$ recorded by a BPM #$j$ at the longitudinal location $z_j$ is proportional to deflection by the dipole corrector $\Delta\theta$

$$\Delta x_j(z_j, t) = \Delta\theta \sqrt{\beta_x(z_j)\beta_{x1}} \sin\varphi_j, \quad (1)$$

where $\beta_{x1}$ and $\beta_x(z_j)$ are Twiss beta-functions in the location of the corrector and BPM, correspondingly, and $\varphi_j$ is the betatron phase advance between the corrector and the BPM.

Let the corrector current change with time in sinusoidal manner with frequency $fc = \frac{\omega}{2\pi}$ producing the deflection amplitude $\theta$, while recording $N_p$ BPM readings with frequency $fb$ at moments $t_k = \frac{k}{fb}$. The set of readings is

$$x_{jk} \equiv \Delta x_j(z_j, t_k) =$$
$$= a0_j \sin \omega t_k = \theta \sqrt{\beta_x(z_j)\beta_{x1}} \sin\varphi_j \sin 2\pi k \frac{fc}{fb}. \quad (2)$$

Application of the Discrete Fourier Transform (DFT) to the data set $\{x_{jk}\}$ delivers the frequency components $c_{jm}$:

$$c_{jm} = \frac{1}{\sqrt{N_p}} \sum_{k=0}^{N_p-1} x_{jk} e^{i\frac{2\pi m}{N_p}k} \quad (3)$$

Since $x_{jk}$ are real numbers, the spectrum is mirror-symmetrical with respect to its center, and below only frequency components $0 \leq m \leq \frac{N_p}{2}$ are considered.

The oscillation frequency and the total number of sampling points can be chosen so that

$$N_p \frac{fc}{fb} = P, \quad (4)$$

where $P$ is an integer indicating the number of full oscillations periods during the measurement. In this case, direct substitution of Eq. (2) and Eq. (4) into Eq. (3) yields only the line at the driving frequency:

$$c_{jm} = \begin{cases} \frac{\sqrt{N_p}}{2} a0_j, & m = P \\ 0, & m \neq P \end{cases} \quad (5)$$

The set of the measured oscillation amplitudes $\{a0_j\}$ for the given deflection defines the orbit response (or "differential trajectory"). Note that DTF returns amplitudes and phases. The phase follows the phase of the exiting signal or, if the positive change of the corrector current results in a negative BPM response, it is shifted by $\pi$. Correspondingly, the differential trajectory component is equal to the amplitude or its negated value.

As soon as the condition of Eq. (4) is fulfilled, the BPM response is a single line in the frequency domain. It allows oscillating simultaneously multiple correctors at different frequencies, extracting the differential trajectory for each corrector from the same BPM signal at the corresponding frequencies [4]. The spectrum will show a peak for each $k$-th corrector at harmonic $P_k$.

The error of measuring the oscillation amplitude is determined by the strength of the noise component at the driving frequency $\sigma_{jL}$. In approximation of low-amplitude white noise around the oscillation frequency, it can be estimated [6] as

$$\sigma_{jP} = \frac{\sigma_{noise}}{\sqrt{2}}, \quad (6)$$

where $\sigma_{noise}$ is the rms value of the recorded frequency components with exclusion of all oscillation frequencies and specific components (e.g. low-frequency drifts or power supply noise) observable without excitation. The signal-to-noise ratio ($S/N = \frac{a0_j}{\sigma_{jP}}$) improves with the number of recorded points as $1/\sqrt{N_p}$, and the differential trajectory can be measured at small perturbations if enough measurement points are recorded. This gives an opportunity to perform such measurements in parallel with regular operation.

## CHOICE OF PARAMETERS

The procedure of choosing the parameters of oscillations is as follows. First, we chose the minimum number of points to record (addressed as a "super period" $M_{psp}$). To keep the minimum measurement time about a minute for the pulse frequency of 5 Hz used in low-power mode of accelerator complex operation, $M_{psp}$ should be ~300. To make sure that the scan goes through different discrete corrector settings every period, $M_{psp}$ can be chosen to be a prime number, and we typically use $M_{psp} = 293$. The number of periods $P_k$ per super period for *k*-th corrector defines the oscillation frequency and needs to be chosen far from drifts (which are typically 0.1 Hz or less) and known noise lines. Usually, we choose the smallest number $P_1 \geq 10$ and assign the others with increment of 1, $P_{k+1} = P_k + 1$.

Another consideration might be important when the number of simultaneously oscillating correctors is large. If deflection by a corrector is nonlinear with settings or the beam passes through a nonlinear element, the BPM spectrum may show a component above noise at double frequency ($2 \cdot P_k$) or at higher harmonics. To avoid interference, it is better to keep $P_k < 2 \cdot P_1$. For example, if the number of simultaneously oscillating parameters is 10, one can use the set of $\{P_k\}$ from 10 to 19. Note that if a nonlinear component is proportional to a product of kicks by two correctors *i* and *j*, (for example, when the beam is moved diagonally), the corresponding response appears at sum $P_i + P_j$ and difference $|P_i - P_j|$ lines.

The upper boundary for the number of periods is placed by the corrector power supplies. In our cases, they are specified for DC mode and respond adequately for settings changing slower than 1 Hz. Therefore, for 15 Hz and 293 points, the number of periods should stay below 20, but can be larger if the machine operates at lower frequency (see example in Fig. 1).

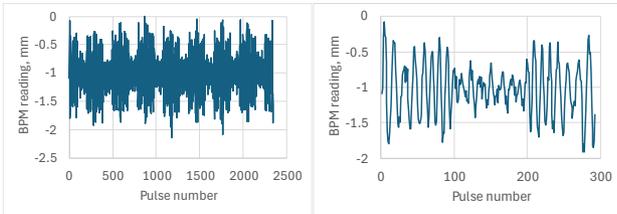

Figure 1: Signal on the last Linac BPM with excitation of 19 same plane correctors upstream. 2 Hz pulse rate. The number of periods per super period is from 20 to 38. Left – full scan (8 super periods), right – expanded view of the first super period.

In practice, first the pattern is run with one super period to verify that it doesn't create extensive losses downstream, and then with a larger number $K_{sp}$ of super periods (we use up to $K_{sp} = 25$) for actual data collection. The S/N ratio improves as $1/\sqrt{K_{sp}}$. The driving frequencies correspond in the Fourier spectrum to peaks at $K_{sp} \cdot P_k$ harmonics. They are separated by $K_{sp} - 1$ points, which makes easier to identify possible problems visually (Fig. 2). For example, if, for some reason, the number of periods per super period is not integer, the response splits into several lines.

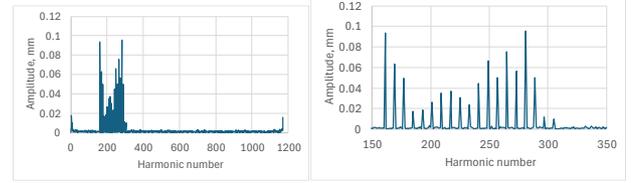

Figure 2: Fourier spectrum of the data shown in Fig.1. Left – full spectrum, right- expanded view over excitation harmonics.

The obvious way to improve S/N is to increase the oscillation amplitude since S/N increases proportionally. The limitations include possible effects on performance of downstream machines or creating excessive losses. One can consider as the figure of merit the area of the phase space covered by the beam during oscillations ("effective emittance") as compared with the emittance of the unperturbed beam. Ref. [6] argues that difference is negligible if the typical oscillation amplitude stays much lower than the beam size. With multiple correctors, the more important factor is the maximum deflection of the beam in sensitive locations. For example, individual oscillations in Fig. 2 result in deviation below 0.1 mm but performing 19 of them simultaneously creates deviations up to 1 mm (Fig. 1). In practice, the amplitudes are chosen empirically based on beam losses.

## REALIZATION AND EXAMPLE

The measurement procedure is used in the Fermilab Linac [7], which accelerates 25 mA, 30 µs pulses of H- ions to 401 MeV. The nominal pulse frequency is 15 Hz, and it may be 1 – 5 Hz for beam studies or in low-power operation mode. The Linac consists of a drift-tube and a side-coupled cavities sections. The latter section is much better instrumented, and the example shown in this report represents measurements there.

The procedure described above was originally tested with a Java data acquisition program and offline analysis in MathCad (see examples in [6]). Presently, it is being implemented as a Python program interacting with Fermilab's control system, ACNET. The user choses the correctors to oscillate with amplitudes, number of points, and number of periods as well as the file with names of parameters to be recorded. After the measurement is performed, the user can check spectra of individual BPMs, and then the table of responses of all BPMs to all correctors is produced. We address this table as Orbit Response Matrix (ORM). The ORM is displayed with color coding to quickly check visually that meaningful components form a triangle (the correctors do not affect BPM readings upstream) and vary with betatron oscillations.

The Python program is still work in progress, and for consistency, all results shown in this paper are from an offline analysis of the same measurement. An example of ORM is presented in Fig.3.

Figure 3: Example of the ORM table showing responses of all horizontal BPMs (in mm) in the high-energy portion of the Linac to all horizontal correctors each oscillating with the amplitude of 0.1 A. The red and green colors highlight values that are, correspondingly, larger than $3\sigma_{jP}$ or smaller than $-3\sigma_{jP}$. Each column corresponds to a corrector, and each row represents a BPM.

At the same time, readbacks and settings of corrector currents are recorded and analysed in the same manner. For comparison with simulations, the ORM columns are normalized by the corresponding corrector amplitudes and expressed in mm/A.

To check the data consistency, one can compare responses to different correctors. In the linear system, any trajectory should be a linear combination of any two other trajectories (downstream of all three correctors). Example of such comparison is shown in Fig. 4. In the case of good quality data, the differences between the trajectory and the best fit should be mostly within the propagated rms measurement error. Note that for Fig. 4, all spectrum of Fig. 2 apart of excitation lines was used to calculate the errors with Eq. (6). While it is a default procedure, in this case existence of the large low-frequency components (drifts) results in an overestimate of the noise at excitation frequencies.

Figure 4: Fitting a differential trajectory to a linear combination of two other trajectories (left) and their difference (right). The bars in the right plot represent the rms measurement errors.

The ORM can be used, for example, for creating a local perturbation or for correcting the beam trajectory over a section of the Linac. For choosing the optimal set of correctors in each situation, it is useful to know how different are their effects. Examples of two scenarios are shown in Fig. 5.

Figure 5: Example of similar (left) and "orthogonal" (right) trajectories. In the left plot, the sign and amplitude of D21TMH trajectory are adjusted for visual comparison.

The first set can be used to amplify deflections, while combination of correctors in the second set allows to adjust both beam shift and angle in a given location independently. The natural way to characterize the similarity or difference would be the betatron phase advance between the correctors. The similar trajectories as in Fig. 5 left are produced by correctors with the phase advance close to $\pi n$, and "very different" as in Fig 5 right are those with the phase advance near $\pi/2+\pi n$ ("orthogonal"). However, presently the accuracy of the Linac lattice reconstruction is not satisfactory to be used for this purpose. Instead, we find that for practical purposes, the difference between two trajectories $\{x_{1i}\}$ and $\{x_{2i}\}$ can be quantified directly from the measured values by the "angle" $\theta_{12}$ between these two vectors introduced as

$$\cos\theta_{12} = \frac{\{x_{1i}\}\cdot\{x_{2i}\}}{|\{x_{1i}\}|\cdot|\{x_{2i}\}|} \equiv \frac{\sum_i x_{1i} x_{2i}}{\sqrt{\sum_i x_{1i}^2 \sum_i x_{2i}^2}}, \qquad (7)$$

where summation is performed over the BPM responses downstream of both correctors. One can show that in the idealized case of a large number of BPMs equally spaced over an integer number of betatron oscillation periods in the lattice with a constant β-function, $\theta_{12}$ is equal to the fractional part of the phase advance between the correctors. While the real case is far from that idealized model, the notion is still helpful (see Ref. [8] as a practical example). It provides a clear distinction between the combinations in in Fig. 5, where $\cos\theta_{12}$ is 1.00 for the left case and 0.04 for the right case. Automated calculation of the cosine is being implemented in the new Python program.

## DISCUSSION

The method described above is applied to measuring the ORM. Note that it is applicable every time when an accurate measurement of the derivative (response) of a "passive" (observable) parameter over an "active" (controllable) parameter is required. We apply it to recording an analogue of ORM in the longitudinal space [9], finding a localized beam loss location [10], and preparing an application of the same technique for centering the beam in optical components. Another avenue to explore is analysis of nonlinear optical elements by looking at higher harmonics of oscillations.